 \documentclass{aastex}
 \shorttitle{New CDM Velocity Distribution}
\shortauthors{Vergados and Owen}

\usepackage{graphicx}% Include figure files
\usepackage{subfig}
\slugcomment{Asymmetric WIMP Distribution in the Eddington Approach.}
\shorttitle{Asymmetric WIMP Velocity Distribution}
\shortauthors{Vergados}
\usepackage{natbib}
\begin{document}
\newcommand{\newc}{\newcommand}
\newc{\ra}{\rightarrow}
\def\vbf{\mbox{\boldmath $\upsilon$}}
\newc{\lra}{\leftrightarrow}
\newc{\beq}{\begin{equation}}
\newc{\eeq}{\end{equation}}
\newc{\barr}{\begin{eqnarray}}
\newc{\earr}{\end{eqnarray}}
%%%%%%%%%%%%%%%%%%%%%%%%%%%%%%%%%%%%%%%%%%%
%\preprint{APS/123-QED}
%\date{\today}
\title{Asymmetric Velocity Distributions from Halo Density Profiles in the 
Eddington Approach}


\author{J.D. Vergados\altaffilmark{1}}
\affil {TEI of Western Macedonia, Kozani, Greece.}
\email{vergados@cc.uoi.gr}

\altaffiltext{1}{Permanent address:Theoretical Physics Division,
 University of Ioannina, Ioannina, Gr 451 10, Greece.}

 \begin{abstract}In the
present paper  we show how obtain the energy  distribution $f(E)$ in our vicinity starting from  WIMP density profiles in a self consistent way by employing the Eddington approach and adding  reasonable angular momentum  dependent terms in the expression of the energy. We then   show how we can  obtain the velocity dispersions and the asymmetry parameter $]beta$ in terms of the  parameters describing the angular momentum dependence. From this expression for $f(E)$ we proceed to construct  an axially symmetric WIMP velocity distributions, which for a gravitationally bound system automatically has an velocity upper bound and is characterized by the  the same asymmetries. This approach is tested and clarified by constructing analytic expressions in a simple model, with adequate structure. We then show how such velocity distributions can be used in determining the event rates, including modulation, both in the standard as well directional WIMP searches. find that some density profiles lead to approximate Maxwell-Boltzmann distributions, which are  automatically defined in a finite domain, i.e. the escape velocity need not be put by hand. The role of such distributions in obtaining the direct WIMP detection rates, including the modulation,  is studied in some detail and, in particular, the role of the asymmetry is explored.
\end{abstract}
%\pacs{95.35.+d, 12.60.Jv}
%\pacs{ 13.15.+g, 14.60Lm, 14.60Bq, 23.40.-s, 95.55.Vj, 12.15.-y}
%%%%%%%%%%%%%%%%%%%%%%%%%%%%%%%%%%%%%%%%%%%%%%%%%%%%%%%%%%%%%%%%%%%%%
%\date{\today}
%\begin{keyword}
Keywords\\
dark matter, galaxies: halos
%\end{keyword}
\maketitle
%\end{keyword}
%\end{frontmatter}
%%%%%%%%%%%%%%%%%%%%%%%%%%%%%%%%%%%%%%%%%%%%%%%%%%%%%%%%%%%%%%%%%%%%%
%%%%%%%%%%%%%%%%%%%%%%%%%%%%%%%%%%%%%%%%%%%%%%%%%%%%%%%%%%%%%%%%%%%%% 
\section{Introduction}
The combined MAXIMA-1 \cite{MAXIMA1},\cite{MAXIMA2},\cite{MAXIMA3}, BOOMERANG \cite{BOOMERANG1},\cite{BOOMERANG2}
DASI \cite{DASI02} and COBE/DMR Cosmic Microwave Background (CMB)
observations \cite{COBE} imply that the Universe is flat
\cite{flat01}
%, $\Omega=1.11\pm0.07$
and that most of the matter in
the Universe is Dark \cite{SPERGEL},  i.e. exotic. These results have been confirmed and improved
by the recent WMAP  \cite{WMAP06} and Planck \cite{PlanckCP13} data. Combining 
the data of these quite precise measurements one finds:
$$\Omega_b=0.0456 \pm 0.0015, \quad \Omega _{\mbox{{\tiny CDM}}}=0.228 \pm 0.013 , \quad \Omega_{\Lambda}= 0.726 \pm 0.015~$$
%$$\Omega_b=0.05, \Omega _{CDM}= 0.25, \Omega_{\Lambda}= 0.70$$ 
(the more  recent Planck data yield a slightly different combination $ \Omega _{\mbox{{\tiny CDM}}}=0.274 \pm 0.020 , \quad \Omega_{\Lambda}= 0.686 \pm 0.020)$. It is worth mentioning that both the WMAP and the Plank observations yield essentially the same value of $\Omega_m h^2$,
% namely  $\Omega_m h^2=0.1423\pm 0.0029$,
  but they differ in the value of $h$, namely $h=0.704\pm0.013$ (WMAP) and $h=0.673\pm0.012$ (Planck).
Since any ``invisible" non exotic component cannot possibly exceed $40\%$ of the above $ \Omega _{\mbox{{\tiny CDM}}}$
~\cite {Benne}, exotic (non baryonic) matter is required and there is room for cold dark matter candidates or WIMPs (Weakly Interacting Massive Particles).\\
Even though there exists firm indirect evidence for a halo of dark matter
in galaxies from the
observed rotational curves, see e.g. the review \cite{UK01}, it is essential to directly
detect such matter in order to 
unravel the nature of the constituents of dark matter. At present there exist many such candidates: the
%provides candidates for the dark matter constituents with the most favored scenario
LSP (Lightest Supersymmetric Particle) \cite{ref2a,ref2b,ref2c,ELLROSZ,Gomez1,Gomez2,ELLFLOR}, technibaryon \cite{Nussinov92,GKS06}, mirror matter\cite{FLV72,Foot11}, Kaluza-Klein models with universal extra dimensions\cite{ST02a,OikVerMou} etc.
  This makes it imperative that we
invest a
maximum effort in attempting to detect dark matter whenever it is
possible. Furthermore such a direct detection will also
 unravel the nature of the constituents of dark matter.\\
 The
 possibility of such detection, however, depends on the nature of the dark matter
 constituents (WIMPs).
 % Here we will assume that such a constituent is the LSP.
  Since the WIMP is expected to be very massive, $m_{\chi} \geq 30 GeV$, and
extremely non relativistic with average kinetic energy $T \approx
50KeV (m_{\chi}/ 100 GeV)$, it can be directly detected
%~\cite{GOODWIT},\cite{KVprd}
 mainly via the recoiling of a nucleus
(A,Z) in elastic scattering. The event rate for such a process can
be computed following a number of steps \cite{JDV06a}.
%\begin{enumerate}
%\item An effective Lagrangian at the elementary particle (quark)
%level obtained in the framework of supersymmetry as described ,
%e.g., in \cite{ref2a,ref2d,JDV96}.
%\item A well defined procedure for transforming the amplitude
%obtained using the previous effective Lagrangian from the quark to
%the nucleon level, i.e. a quark model for the nucleon. This step
%is not trivial, since the obtained results depend crucially on the
%content of the nucleon in quarks other than u and d. This is
%particularly true for the scalar couplings, which are proportional
%to the quark masses~\cite{Dree1}, \cite{Chen2}, \cite{JDV06a} as well as the
%isoscalar axial coupling \cite{JELLIS93a,JDV06a}.
%\item Knowledge of the relevant nuclear matrix elements
%\cite{Ressa,Ressb,DIVA00,KVprd}, obtained with as reliable as possible many
%body nuclear wave functions. Fortunately in the case of the scalar
%coupling, which is viewed as the most important, the situation is
%a bit simpler, since  then one  needs only the nuclear form
%factor.
%\item 
In the present work we will focus on the WIMP density in our vicinity and its velocity distribution.
% Since the
%essential input here comes from the rotational curves,  dark matter candidates other than the
%LSP (neutralino) are also characterized
%by similar parameters.
% In most calculations performed up to now
%one employs a Maxwell-Boltzmann velocity distribution with an
%upper velocity cut off put in by hand at the escape velocity $\upsilon_{esc}=2.84 \upsilon_0\simeq 620$ km/s.
%\end{enumerate}

In the past various velocity distributions have been considered.
The one most used is the isothermal Maxwell-Boltzmann velocity
distribution with $<\upsilon ^2>=(3/2)\upsilon_0^2$ where
$\upsilon_0$ is the velocity of the sun around the galaxy, i.e.
$220~km/s$. Extensions of this M-B distribution were also
considered, in particular those that were axially symmetric with
enhanced dispersion in the galactocentric direction
 \cite {Druka,Drukb,Verg00,evans00}. In all such
distributions an upper cutoff $\upsilon_{esc}=2.84\upsilon_0$ was introduced
by hand, in the range obtained by Kochanek\cite{Kochanek95}. In a different approach  Tsallis type functions, derived from simulations of dark matter densities were employed, see e.g. recent calculations \cite{VerHanH} and references there in .

Non isothermal models have also been considered. Among those one should
mention the late infall of dark matter into the galaxy, i.e caustic rings
 \cite{SIKIVI1,SIKIVI2,Verg01,Green,Gelmini}, dark matter orbiting the
 Sun \cite{Krauss}, Sagittarius dark matter \cite{GREEN02}.

The correct approach in our view is to consider the Eddington
proposal \cite{EDDIN}, i.e. to obtain both the density and the
velocity distribution from a mass distribution, which depends both
on the velocity and the gravitational potential. Our motivation in
using Eddington \cite{EDDIN} approach to describing the density of
dark matter is found, of course, in his success in describing the
density of stars in globular clusters. Since this approach
adequately describes the distribution of stars in a globular
cluster in which the main interaction is gravitational and because
of its generality , we see no reason why such an approach should
not be applicable to dark matter that also interacts
gravitationally.
It seems, therefore, not surprising that this  approach has been
used by Merritt \cite{MERRITT} and  applied to dark matter by
Ullio and Kamionkowski\cite{ULLIO} and  by us
\cite{OWVER,VEROW06}.

 It is the purpose of the present paper to extend the previous work obtain a
dark matter velocity distribution, which need not be spherically, but they may originate from density profiles that are spherically symmetric.We have constructed a one-parameter family of self-consistent star clusters that are spherically
symmetric but anisotropic in velocity space. These were computed modifying the distribution (DF) by including suitable angular momentum factors along the lines suggested by Wojtak {\it et al} \cite{Asym08} and more recently by Fornasa and Green \cite{FornGreen13}. Also a one-parameter family of self-consistent star clusters that are spherically
symmetric was shown to be  anisotropic in velocity space \cite{NguPed13} (see also \cite{AgPedRM11}). The model was constructed first in the Newtonian limit and then after the first post-Newtonian corrections were computed. To clarify some of the issues involved in this approach, we will concentrate on some cases amenable to analytic solutions like the celebrated Plummer solution \cite{PLUMMER}. We will show how this method can be used in dark matter searches and leave  the case of realistic calculations for a future publication.

\section{The Dark Matter Distribution in the Context of  the Eddington approach}
The introduction the matter distribution can be given\cite{VEROW06} as follows
\beq dM=2\pi~f(\Phi({\bf r}),\upsilon_r,\upsilon_t)~dx~dy~dz
~\upsilon_t~d\upsilon_t~d\upsilon_r \label{distr.1} \eeq where the
function $f$ the distribution function, which depends on ${\bf r}$
through the potential $\Phi({\bf r})$ and the tangential and
radial velocities $\upsilon_t$ and $\upsilon_r$. 
We will limit ourselves in spherically symmetric systems.
%In general the
%distribution function is not symmetric. In the above expression we
%assumed that it is only axially symmetric, with the two tangential
%components being equal.
 Then the density of matter $\rho(|r|)$
satisfies the equation:
  \beq
d\rho=2\pi~f(\Phi(|{\bf r}|
),\upsilon_r,\upsilon_t)~~\upsilon_t~d\upsilon_t~d\upsilon_r
\label{distr.2} \eeq
\subsection{The distribution is a function of the total energy only}
%\begin{itemize}
%\item
 The energy is given by $E=\Phi(r)+\frac{\upsilon^2}{2}$. Then
\beq
\rho(r)=4 \pi\int f(\Phi(r)+\frac{\upsilon^2}{2})\upsilon^2 d\upsilon=4 \pi\int_{\Phi}^0 {f(E)}{\sqrt{2(E-\Phi)}}dE
\label{Eq:rhoE}
\eeq
This is an integral equation of the Abel type. It can be inverted to yield:
\begin{equation}
f(E)=\frac{\sqrt{2}}{4 \pi^2}\frac{d}{dE} \int_E^0 \frac{d \Phi}{\sqrt{\Phi-E}}\frac{d \rho}{d \Phi}
\end{equation}
The above equation can be rewritten as:
\begin{equation}
f(E)=\frac{1}{2 \sqrt{2} \pi ^2} \left[ \int_E^0 \frac{d\Phi}{\sqrt{\Phi-E}} \frac{d^2 \rho}{d \Phi ^2}-
\frac{1}{\sqrt{-E}} \frac{d \rho}{d \Phi}|_{\Phi=0} \right ]
\label{Eq:sol2}
\end{equation}
 In order to proceed it is necessary to know the density as a function of the potential. In practice only in few cases this can be done analytically. This, however, is not a problem, since this function can be given parametrically by the set $(\rho(r),\Phi(r))$ with the position $r$ as a parameter. The potential $\Phi(r)$ for a given density $\rho(r)$ is obtained by solving Poisson's equation.\\
Once the function f(E) is known we can obtain the needed velocity distribution $f_{r_s}(\upsilon)$ in our vicinity ($r=r_s$) by writing:
\beq
f_{r_s}(\upsilon^{\prime})={\cal N}f(\Phi(r)|_{r=r_s}+\frac{\upsilon^{\prime 2}}{2})
\eeq
where ${\cal N}$ is a normalization factor.
\subsection{ Angular momentum dependent terms}
The presence of such terms can introduce asymmetries in the velocity dispersions.
%We suppose now that there are asymmetries in the density and the velocity distributions.\\
In such a a reasonable model \cite{Asym08} we get:
\beq
\rho({\bf r})=\int\int \int f(E) \left (1+\frac{L^2}{2 L^2_0}\right )^{-\beta_{\infty}+\beta_0} L^{- 2 \beta_0} d^3 \vbf. \def\vbf{\mbox{\boldmath $\upsilon$}}
\label{Eq:rhogen}
\eeq
i.e. by introducing three new parameters. Introducing the new parameters $L$ and $E$ in terms of $\upsilon_t$ and  $\upsilon_r$ via:
$$\upsilon_t=\frac{L}{r},\,\upsilon_r=\sqrt{2(E-\Phi)-\frac{L^2}{r^2}} \mbox{ or } \upsilon_t=\frac{L_0}{r}\sqrt{2 \lambda},\,\upsilon_r=\sqrt{2}\frac{L_0}{r}\sqrt{x-\lambda},\,\lambda=\frac{L^2}{2 L_0^2}$$
we can perform the integration in cylindrical coordinates and get:
%Noting that $\upsilon_t$, the tangential component of the velocity is related to the angular momentum via $\upsilon_t=L/r$,  the integration over the velocities can performed to yield:
\beq
\rho({\bf r})=2^{1/2-\beta_0}L_0^{1-2 \beta_0} \frac{\pi}{r}\int^0_{\Phi}f(E) dE \int_0^x \frac{\lambda ^{-\beta_0} (\lambda +1)^{-\beta_{\infty}+
   \beta_0}}{\sqrt{x-\lambda }}d \lambda
   \label{rhoas}
\eeq 
In the above expressions $x=(r^2/L_0^2)(\Phi-E)$. \\Before proceeding further we prefer to write the above formula in terms of dimensionless variables $\Phi=\Phi_0 \xi$, $\rho=\rho_0 \eta$, $E=\Phi_0 \epsilon$ and $f(E)=\rho_0 \Phi_0^{-3/2}\tilde {f} (\epsilon)$. Thus the last equation becomes:
\beq
\eta=2^{1/2-\beta_0}L_0^{-2 \beta_0}\frac{1}{\sqrt{a}} \pi \int^0_{\xi}\tilde{f}(\epsilon ) d\epsilon  \int_0^x \frac{\lambda ^{-\beta_0} (\lambda +1)^{-\beta_{\infty}+
   \beta_0}}{\sqrt{x-\lambda }}d \lambda
   \label{rhoas1}
\eeq 
with $a=\frac{r^2\Phi_0^2}{L_0^2}$ and $x=a (\xi-\epsilon)$.
 
 The second integral can be done analytically to yield:
$$\frac{\sqrt{\pi } x^{\frac{1}{2}-\beta_0} \Gamma \left(1-\beta
   _{0 }\right)}{\Gamma \left(\frac{3}{2}-\beta _{0 }\right)}~ _2F_1(1-\beta_0,-\beta_0+\beta_{\infty},3/2-\beta_0,-x),$$
 with  $_2F_1 $ the usual hypergeometric function.
 Then Eq. (\ref{rhoas}) becomes:
 \barr
 \eta&=&2^{1/2-\beta_0}L_0^{-2 \beta_0}\frac{1}{\sqrt{a}} \pi\frac{\sqrt{\pi }  \Gamma \left(1-\beta
   _{0 }\right)}{\Gamma \left(\frac{3}{2}-\beta _{0 }\right)}\int_{\xi}^0 \tilde{f}(\epsilon)d\epsilon x^{\frac{1}{2}-\beta_0}\nonumber\\ 
   & &~_2F_1(1-\beta_0,-\beta_0+\beta_{\infty},3/2-\beta_0,-x)
 \earr
In the limit in which $\beta_0->0,\, L_0->\infty$ the last expression is reduced to Eq. (\ref{Eq:rhoE}).\\
Eq (\ref{Eq:rhogen}) allows the calculation of moments of the velocity. In particular following the procedure of 
 \cite{Asym08} one finds:
 \barr
 \prec\upsilon_t^2\succ&=&2 \left (  \frac{L_0}{r} \right )^2(2-\beta_0)\nonumber\\
 && \frac{\int_{\xi}^0 \tilde{ f}(\epsilon)d\epsilon x^{3/2-\beta_0}~_2F_1(2-\beta_0,-\beta_0+\beta_{\infty},5/2-\beta_0,-x)}{\int_{\xi}^0 \tilde{ f}(\epsilon)d\epsilon x^{1/2-\beta_0}~_2F_1(1-\beta_0,-\beta_0+\beta_{\infty},3/2-\beta_0,-x)}
 \earr
 \barr
 \prec\upsilon_r^2\succ&=& \left (  \frac{L_0}{r} \right )^2(1-\beta_0)\nonumber\\
 && \frac{\int_{\xi}^0 \tilde{ f}(\epsilon)d\epsilon x^{3/2-\beta_0} ~_2F_1(1-\beta_0,-\beta_0+\beta_{\infty},5/2-\beta_0,-x)}{\int_{\xi}^0 \tilde{ f}(\epsilon)d\epsilon x^{1/2-\beta_0}~_2F_1(1-\beta_0,-\beta_0+\beta_{\infty},3/2-\beta_0,-x)}
 \earr
 The extra factor of 2 in the case of the tangential velocity can be understood, since there exist two such components. The moments of the velocity are, of course, functions of the three parameters of the model. The model clearly can accommodate asymmetries in the velocity dispersion, even if the density is spherically symmetric.\\ 
 Eq. (\ref{rhoas}) can be inverted to yield the distribution function $\tilde{ f}(\epsilon) $, even though this is technically more complicated than in the standard Eddington approach without the angular momentum factors. Given the function $\tilde{ f}(\epsilon) $ we define the quantities:
 \beq
 \Lambda_t=(2-\beta_0) \int_{\xi}^0 \tilde{ f}(\epsilon)d\epsilon  ~_2F_1(2-\beta_0,-\beta_0+\beta_{\infty},5/2-\beta_0,-x),
 \eeq
 \beq
 \Lambda_r=(1-\beta_0) \int_{\xi}^0 \tilde{ f}(\epsilon)d\epsilon  ~_2F_1(1-\beta_0,-\beta_0+\beta_{\infty},5/2-\beta_0,-x)
 \eeq
 Then the asymmetry parameter $\beta$ defined by:
 \beq
 \beta=1-\frac{\prec\upsilon_t^2\succ}{2 \prec\upsilon_r^2\succ},
 \eeq
 is given by:
  \beq
 \beta=1-\frac{\Lambda_t}{\Lambda_2}.
 \eeq
 The axially symmetric velocity distribution, with respect to the center of the galaxy, is thus obtained from $f(E)$ as described in the Appendix  below.
% \beq
% f_{r_s}(\vbf^{\prime})=\frac{{\cal N}}{(1-\beta)}f(\Phi(r_s)+\frac{1}{2}\frac{1}{1-\beta}\left(\upsilon^{\prime 2}-\beta \upsilon_r^{\prime 2}\right )
% \eeq
% where $\Phi(r_s)$ is the value of the potential in our vicinity and ${\cal N}$ a suitable normalization constant.\\
  Clearly for a given matter density profile , both the  distribution function $\tilde{ f}(\epsilon) $ as well as the integrals $\Lambda_t$ and $\Lambda_r$ are functions of the parameters $r_s$ $\beta_0$ $\beta_{\infty}$ and $L_0$. So is the asymmetry parameter $\beta$. The above equations get simplified in the following cases:
 \begin{enumerate}
 \item In the limit in which $\beta_0=0$ and $\beta_{\infty}=-1$. Then
 \beq
 \eta=4 \pi \int_{\xi}^0 \tilde{ f}(\epsilon)d\epsilon \sqrt{2(\epsilon-\xi)}\left (1+\frac{2}{3} a (\epsilon-\xi) \right ),\,a=\frac{r^2\Phi_0}{L^2_0}
 \label{Eq:ModDis}
 \eeq
  \beq
 \prec\upsilon_t^2\succ=\frac{2}{15}\frac{L_0^2}{r^2}
  \frac{\int_{\xi}^0 \tilde{ f}(\epsilon)d\epsilon  (\epsilon-\xi)^{3/2}(5+4 a(\epsilon-\xi))}{\int_{\xi}^0 \tilde{ f}(\epsilon)d\epsilon \sqrt{\epsilon-\xi}(1+(2/3) a(\epsilon-\xi))}
 \eeq
  \beq
 \prec\upsilon_r^2\succ=\frac{1}{15}\frac{L_0^2}{r^2}
  \frac{\int_{\xi}^0 \tilde{ f}(\epsilon)d\epsilon  (\epsilon-\xi)^{3/2}(5+2 a(\epsilon-\xi))}{\int_{\xi}^0 \tilde{ f}(\epsilon)d\epsilon \sqrt{\epsilon-\xi}(1+(2/3)a(\epsilon-\xi))}
 \eeq
 \beq
 \beta=1-\frac{\int_{\xi}^0  \tilde{ f}(\epsilon)d\epsilon(\epsilon-\xi)^{3/2}(5+4 a(\epsilon-\xi))}{\int_{\xi}^0 \tilde{ f}(\epsilon)d\epsilon (\epsilon-\xi)^{3/2}(5+2 a(\epsilon-\xi))}
 \eeq
 \item $\beta_{\infty}=1$, $\beta_0=0$.\\
  In this case:
 $$\frac{1}{\sqrt{a}} x^{\frac{1}{2}-\beta_0} 
  ~_2F_1(1-\beta_0,-\beta_0+\beta_{\infty},3/2-\beta_0,-x)\rightarrow \frac{1}{\sqrt{a}} \frac{\sinh^{-1}(\sqrt{x})}{\sqrt{1+x}}$$ 
  This function is very complicated to handle. Note however that for sufficiently small values of $a$ one finds that the above expression for $x=a(\epsilon-\xi)$  is reduced to:
  $$ 2 \sqrt{\epsilon-\xi}\left (1-\frac{2}{3} a (\epsilon-\xi)\right )$$
  We thus recover the previous formula with just a change of sign in $a$.
   The corresponding expressions for the velocity dispersions become:
  $$ \Lambda_t\Leftrightarrow 2\left (\sqrt{x}-\frac{\sinh^{-1}(\sqrt{x})}{\sqrt{1+x}}  \right ),\, \Lambda_r\Leftrightarrow 4\left (-\sqrt{x}+\sqrt{1+x}\sinh^{-1}(\sqrt{x})  \right )$$
  In the limit of small $a$ we again recover the previous expressions with $a\rightarrow-a$.
 \item The case of $L>>L_0$.\\
 In this case the integral equation:
 \beq
 \eta=\pi \sqrt{2 \pi}a^{-\beta_{\infty}}\frac{\Gamma(1-\beta_{\infty})}{\Gamma(3/2-\beta_{\infty})}\int_\xi^0(\epsilon-\xi)^{1/2-\beta_{\infty}}\tilde{f}(\epsilon)d \epsilon
 \eeq 
 can be solved exactly(see Appendix below) to yield:
  \barr
 \tilde{f}(\epsilon)&=&\frac{a^{\beta_{\infty}}}{\pi^2 \sqrt{2 \pi}}\frac{\Gamma(3/2-\beta_{\infty})}{\Gamma(1-\beta_{\infty})}\frac{\sin{(\pi(1/2-\beta_{\infty}))}}{(1/2-\beta_{\infty})}\nonumber\\&& \frac{d}{d \epsilon}\int_\epsilon ^0(\xi-\epsilon)^{-1/2+\beta_{\infty}}\frac{d \eta(\xi)}{d \xi} d \xi,
 \earr
  provided $\eta(0)=0$. In this case, however, we find that 
  $$
 \beta=1-\frac{\Lambda_t}{\Lambda_r}=1-\frac{\Gamma(2-\beta_{\infty})}{\Gamma(1-\beta_{\infty})}=1-\beta_{\infty},\,\beta_{\infty}<1
  $$
  regardless of the velocity distribution.
  \end{enumerate}
 \section{ Asymmetries in the velocity distribution}
Proceeding as above we get the function $f_{(\beta_{\infty},\beta_0,L_0)}(E)$. We then proceed to construct a velocity distribution, which is characterized by the same asymmetry in velocity dispersion along  lines similar to those previously adopted 
% So the asymmetry in the velocity distribution is contained in writing properly the energy as a function of the velocity. This achieved by including
%an additional kinetic term associated with an angular momentum 
\cite{bookBT87}, i.e. by considering  models of the Osipkov-Merritt type \cite{OSIPKOV79,MERRITT,MERRITT85b}. Thus the velocity distribution in our vicinity $(r=r_s)$ is written as:
\beq
f_{r_s}(\vbf)={\cal N}(1+\alpha_s)f_{0,0,\infty}\left (\Phi(r_rs)+\frac{\upsilon^{\prime 2}_r}{2}+\left (1+\alpha_s\right )\frac{\upsilon^{\prime 2}_t}{2}\right)
\eeq
%\barr
%Q\equiv E&+&\alpha_s \frac{J^2}{2 r^2}=\Phi(r)+\frac{\upsilon^2}{2}+\alpha_s \frac{|r\times v|^2}{2 r^2}=\Phi(r)+\frac{\upsilon^2}{2}+\alpha_s \frac{\upsilon^2_t}{2}
%\nonumber\\
%&=&\Phi(r)+\frac{\upsilon^2_r}{2}+\left (1+\alpha_s\right )\frac{\upsilon^2_t}{2}
%\earr
where $\upsilon^{\prime}_r$ and $\upsilon^{\prime}_t$ are the radial, i.e. outwards from the center of the galaxy, and the tangential components of the velocity, with respect to the center of the galaxy.
%the polar axis is chosen in the radial direction
The parameter   $\alpha_s=\beta/(1-\beta)$ can be determined by calculating the moments of the velocity as above, i.e. it is a function of the parameters $L_0,\,\beta_0$ and $\beta_{\infty}$. since these parameters are usually treated as phenomenological parameters, we will treat $\beta$ phenomenologically.
We note that this function is only axially symmetric and the normalization constant {\cal N} is a normalization constant, the same as in the case of $\alpha_s=0$. The isotropic case follows as a special case in the limit $\alpha_s\rightarrow0$.

 The characteristic feature of this approach is that the velocity distribution automatically vanishes outside a given region specified by a cut off velocity $\upsilon_m$, given by $\upsilon_m=\sqrt{2 |\Phi(r_s)|}$.
% \end{itemize}
 \section{A simple test density profile}
 Before proceeding further  we will examine a simple  model, amenable to analytic solution, i.e. the famous Plummer solution \cite{PLUMMER}  and leave the case of realistic density profiles, like, e.g., those often employed   \cite{NFW}, \cite{ULLIO}, \cite{VEROW06} for a future publication. It is well known that a spherical density distribution   \cite{PLUMMER} of the type
 \beq
 \eta=\frac{\rho(x)}{\rho_0}= \frac{1}{(1+x^2/3)^{5/2}},\quad x=\frac{r}{a},
 \eeq
 which is sometimes used as an  ordinary matter profile, leads to a potential of the form 
 \beq
 \xi=\frac{\Phi(x)}{\Phi_0} =-\frac{1}{(1+x^2/3)^{1/2}},\, \Phi_0=4 \pi G_N a^2 \rho_0
 \eeq
 It  is interesting to  remark that the Plummer solution naturally arises in  a model involving  self-consistent star clusters   studied
the Newtonian limit as well as after the first post-Newtonian corrections were computed \cite{NguPed13}.
 
 From these we obtain the desired relation:
 \beq
 \eta(\xi)=-\xi^5, \mbox{ with } \eta^{\prime \prime}(\xi)=-20 \xi^3, \left . \eta(\xi) \right | _{\xi=0}=0, \left . \frac{d \eta}{d \xi} \right | _{\xi=0}=0
 \eeq
 Then the solution to Eq. \ref{Eq:ModDis} is given by 
 \barr
 \tilde {f}(x)&=&\frac{16 e^{-a x}}{a^{9/2} \pi  x}\nonumber\\
 && e^{a x} \left(\sqrt{a} \sqrt{x} (2 a x (2 a x (2 a
   x-5)+15)-15)+15 \sqrt{a x}\right)\nonumber\\
   &-&15 a \sqrt{\pi }
   x \mbox{erfi}\left(\sqrt{a x}\right),\quad x=-\epsilon
 \earr
 This leads to a velocity distribution
 \beq
 f_{\xi(xs)}(y)=\tilde {f}(\xi(xs)-y^2/2)
    \eeq
  where $\xi(xs) $ is the value of the potential in our vicinity. In our simple model $\xi(xs)\approx \sqrt{3}/2$.  We also used a larger value $\xi(xs)=10$. 
  \begin{enumerate}
  \item The choice $a>0$\\
  The obtained velocity distribution  properly normalized is exhibited in Fig. \ref{Fig:toyfv}. We notice that the dependence on $a$ is very mild.\\
     \begin{figure}[!ht]
 \begin{center}
 \subfloat[]
{
\rotatebox{90}{\hspace{-0.0cm}{$4 \pi y^2 f_{xs}(y)\rightarrow$}}
%\rotatebox{90}{\hspace{-0.0cm} {$\Phi(x)\longrightarrow 4 \pi G_Na^2 \rho_0$}}
%\includegraphics[scale=0.6]{NFWlogrhophi.eps}
\includegraphics[scale=0.7]{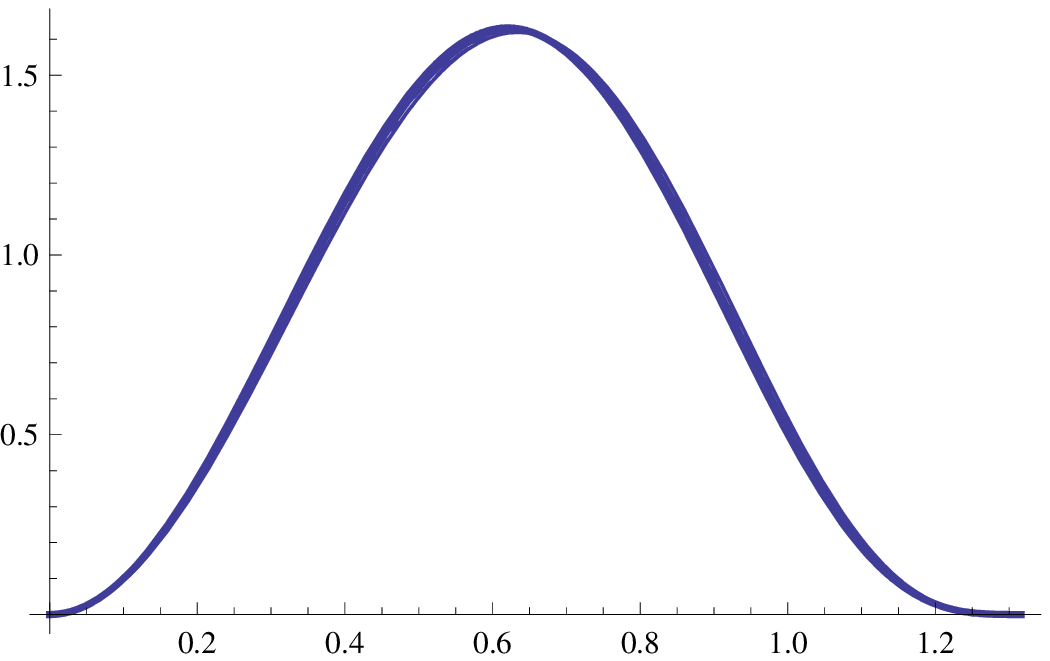}
}
\subfloat[]
{
\rotatebox{90}{\hspace{-0.0cm}{$4 \pi y^2 f_{xs}(y)\rightarrow$}}
%\rotatebox{90}{\hspace{-0.0cm} {$\Phi(x)\longrightarrow 4 \pi G_Na^2 \rho_0$}}
%\includegraphics[scale=0.6]{NFWlogrhophi.eps}
\includegraphics[scale=0.7]{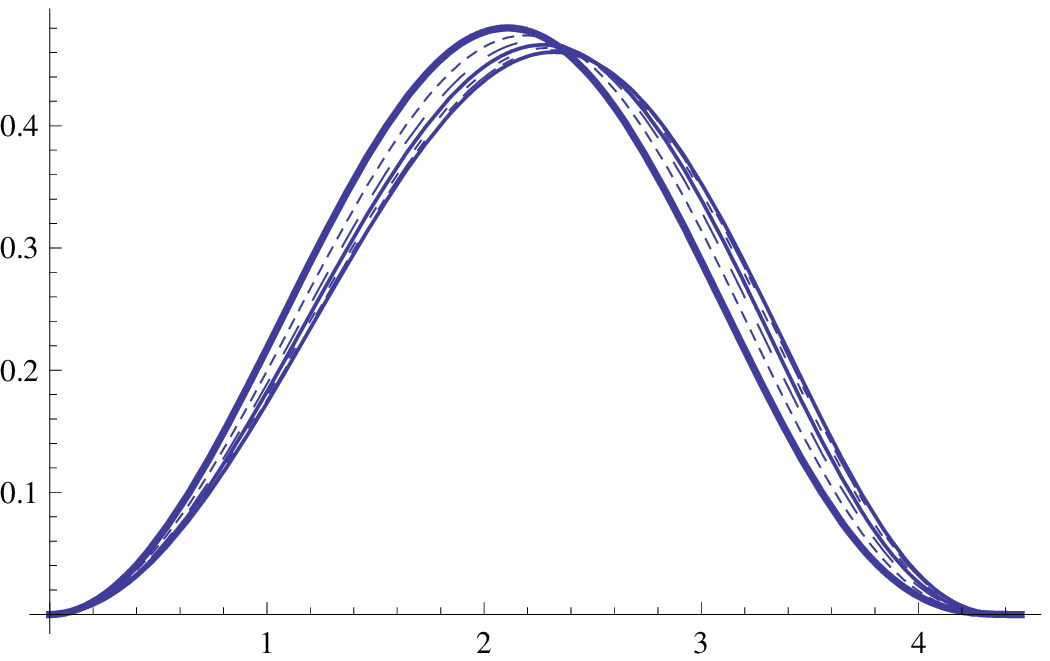}
}\\
{\hspace{-0.0cm} {$y\longrightarrow $}}\\
 \caption{We show the properly normalized velocity distribution obtained in our simple model for various values of $a$ for the value $\xi(xs)=\sqrt{3}/2$ (a) and a larger, perhaps more realistic, value $\xi(xs)=10$ (b). The obtained velocity distribution depends mildly on $a$.}
 \label{Fig:toyfv}
  \end{center}
  \end{figure}
  We next compute the asymmetry parameter $\beta=1-\Lambda_t/\Lambda_r$ as a function the potential $\xi$ for various values of $a$. This is exhibited in Fig. \ref{Fig:toybeta}. The asymmetry is negative, opposite to what is commonly believed, see e.g. 
  \cite {Druka,Drukb,Verg00,evans00},
   \cite{HANSEN06a},\cite{VerHanH}, i.e.  it does not lead to enhanced dispersion in the galactocentric direction,  regardless of the values of  $\xi$. Thus the positive values of $a$ are not acceptable, i.e. the choice  $\beta_{\infty}=-1$, $\beta_{0}=0$ is not physically acceptable.
       \begin{figure}[!ht]
 \begin{center}
 
\rotatebox{90}{\hspace{-0.0cm}{$\beta\rightarrow$}}
%\rotatebox{90}{\hspace{-0.0cm} {$\Phi(x)\longrightarrow 4 \pi G_Na^2 \rho_0$}}
%\includegraphics[scale=0.6]{NFWlogrhophi.eps}
\includegraphics[scale=1.0]{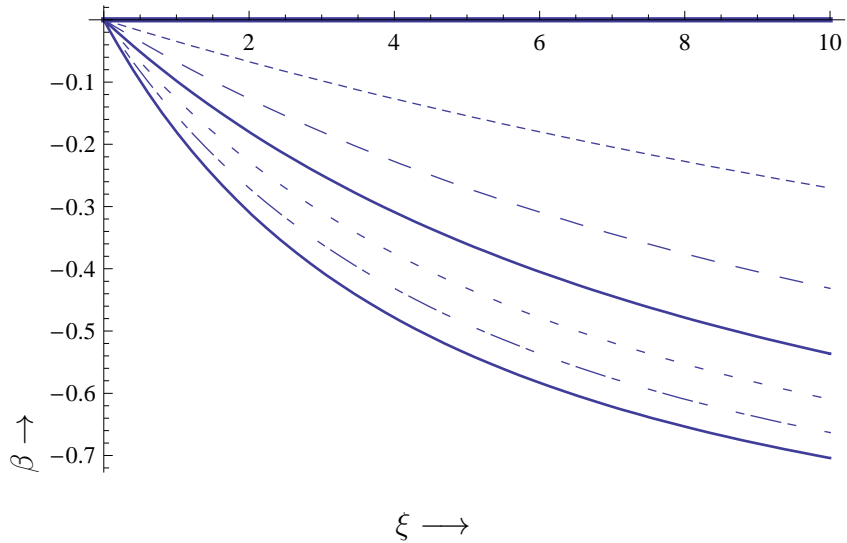}
\\
{\hspace{-0.0cm} {$\xi\longrightarrow $}}\\
 \caption{The asymmetry parameter $\beta=\Lambda_t/\Lambda_r$ as a function of $\xi$ for values of $a=0,0.25,0.50,0.75,1.0,1.25,1.50$ increasing downwards.}
 \label{Fig:toybeta}
  \end{center}
  \end{figure}
   \item The choice $\beta_{\infty}=1$, $\beta_{0}=0$.\\
   In this case we will explore the regime of negative  absolutely small values  of $a$
   The velocity distribution obtained is exhibited in Fig.  \ref{Fig:toyfv1}  
     \begin{figure}[!ht]
 \begin{center}
 \subfloat[]
{
\rotatebox{90}{\hspace{-0.0cm}{$4 \pi y^2 f_{xs}(y)\rightarrow$}}
%\rotatebox{90}{\hspace{-0.0cm} {$\Phi(x)\longrightarrow 4 \pi G_Na^2 \rho_0$}}
%\includegraphics[scale=0.6]{NFWlogrhophi.eps}
\includegraphics[scale=0.5]{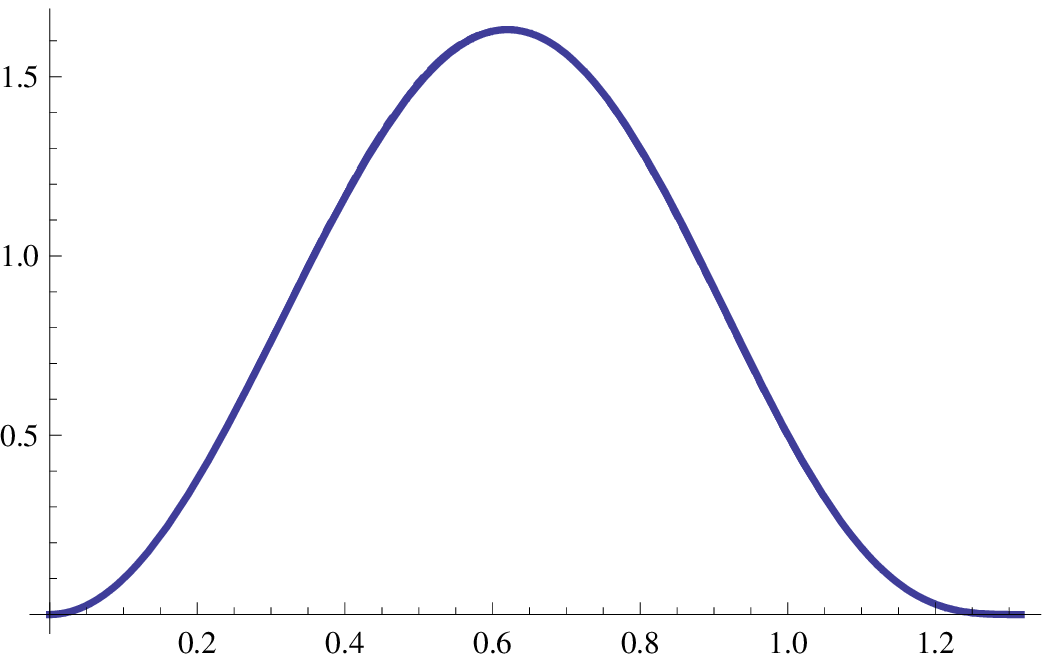}
}
\subfloat[]
{
\rotatebox{90}{\hspace{-0.0cm}{$4 \pi y^2 f_{xs}(y)\rightarrow$}}
%\rotatebox{90}{\hspace{-0.0cm} {$\Phi(x)\longrightarrow 4 \pi G_Na^2 \rho_0$}}
%\includegraphics[scale=0.6]{NFWlogrhophi.eps}
\includegraphics[scale=0.8]{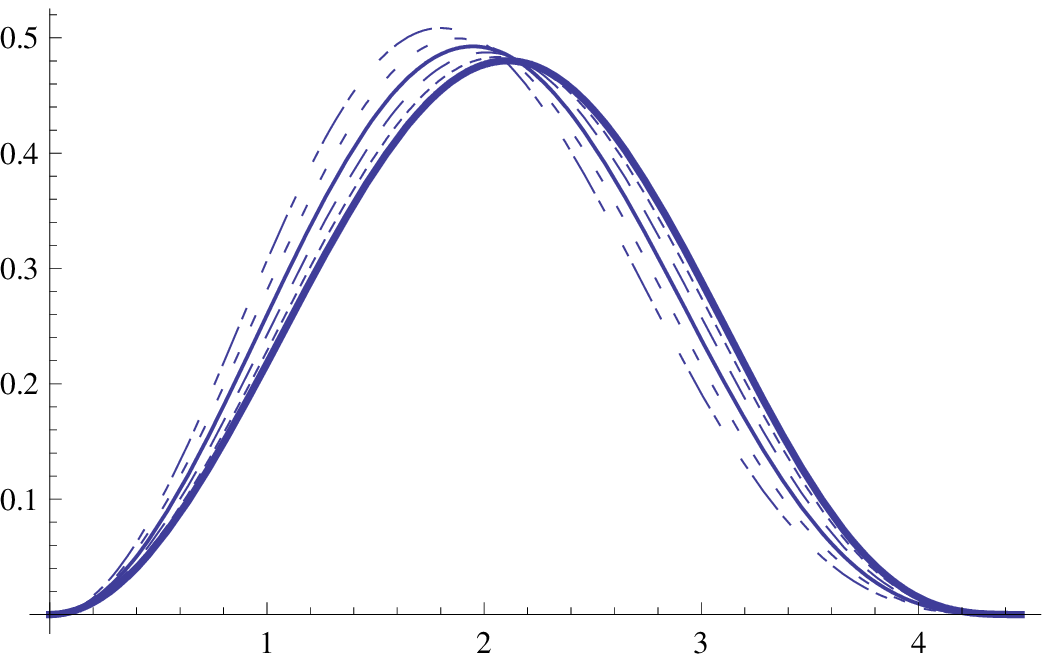}
}\\
{\hspace{-0.0cm} {$y\longrightarrow $}}\\
 \caption{We show the properly normalized velocity distribution obtained in our simple model for negative values of $a$, i.e ${a=0,-0.1,-0.2,-0.3,-0.4,-0.5}$ for the value $\xi(xs)=\sqrt{3}/2$ (a) and a larger, perhaps more realistic, value $\xi(xs)=10$ (b). The obtained velocity distribution depends mildly on $a$ in (a) and it is noticeable in (b. In the plots a is increasing from left to right.}
 \label{Fig:toyfv1}
  \end{center}
  \end{figure}
  
  \begin{figure}
  \begin{center}
\rotatebox{90}{\hspace{-0.0cm}{$\beta\rightarrow$}}
%\rotatebox{90}{\hspace{-0.0cm} {$\Phi(x)\longrightarrow 4 \pi G_Na^2 \rho_0$}}
%\includegraphics[scale=0.6]{NFWlogrhophi.eps}
\includegraphics[scale=1.0]{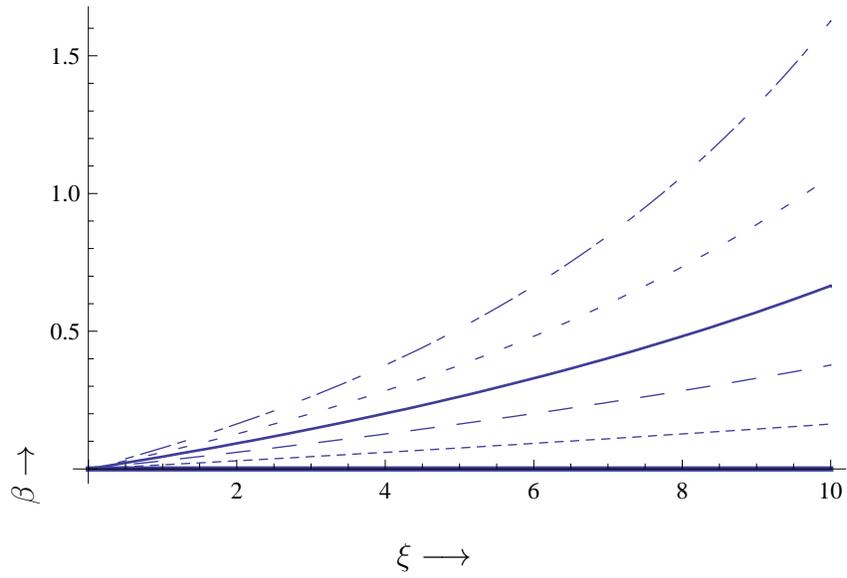}
\\
{\hspace{-0.0cm} {$\xi\longrightarrow $}}\\
 \caption{The asymmetry parameter $\beta=\Lambda_t/\Lambda_r$ as a function of $\xi$ for values of $a$ the same as in Fig.  \ref{Fig:toyfv1}. In the plots $a$ is increasing upwards.}
 \label{Fig:toybeta1}
  \end{center}
  \end{figure}
 
  \begin{figure}
  \begin{center}
\rotatebox{90}{\hspace{-0.0cm}{$\beta\rightarrow$}}
%\rotatebox{90}{\hspace{-0.0cm} {$\Phi(x)\longrightarrow 4 \pi G_Na^2 \rho_0$}}
%\includegraphics[scale=0.6]{NFWlogrhophi.eps}
\includegraphics[scale=1.0]{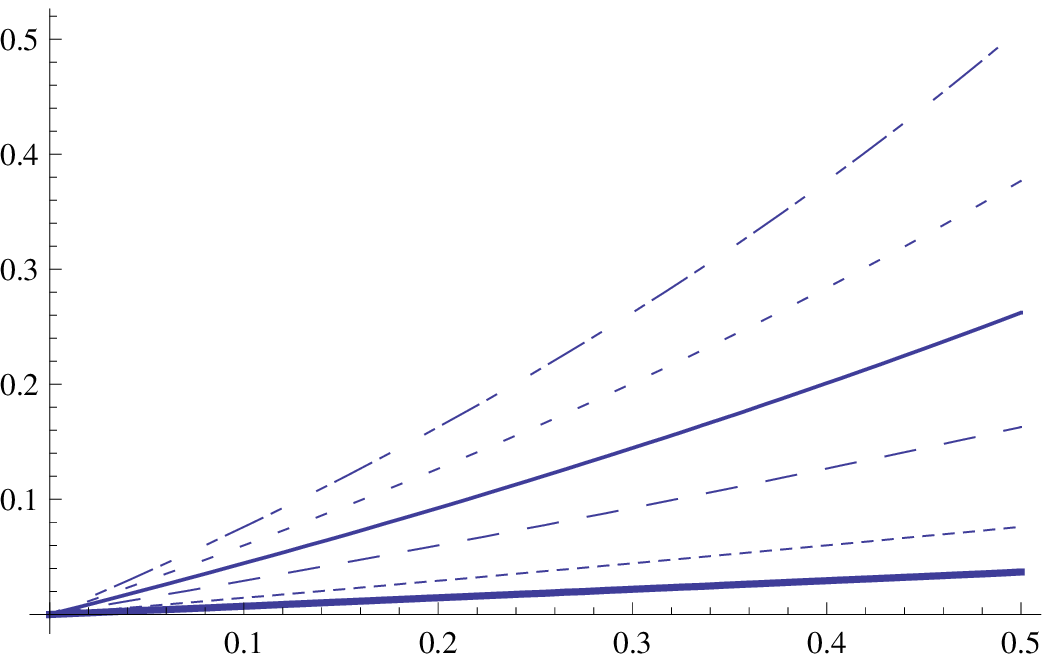}
\\
{\hspace{-0.0cm} {$-a\longrightarrow $}}\\
 \caption{The asymmetry parameter $\beta=\Lambda_t/\Lambda_r$ as a function of $a$ for values of $\xi (r_s)=(1,2,4,6,8,10)/2$. In the plots $\xi(r_s)$ is increasing upwards. Note that on the x-axes the opposite of $a$ is indicated.}
 \label{Fig:betavsa}
  \end{center}
  \end{figure}
   \end{enumerate}
   \section{The velocity distribution in WIMP searches}
\begin{figure}
 \begin{center}
\rotatebox{90}{\hspace{-0.0cm}{$4 \pi y^{\prime2} f(\beta,y^{\prime})\rightarrow$}}
%\rotatebox{90}{\hspace{-0.0cm} {$\Phi(x)\longrightarrow 4 \pi G_Na^2 \rho_0$}}
%\includegraphics[scale=0.6]{NFWlogrhophi.eps}
\includegraphics[scale=0.7]{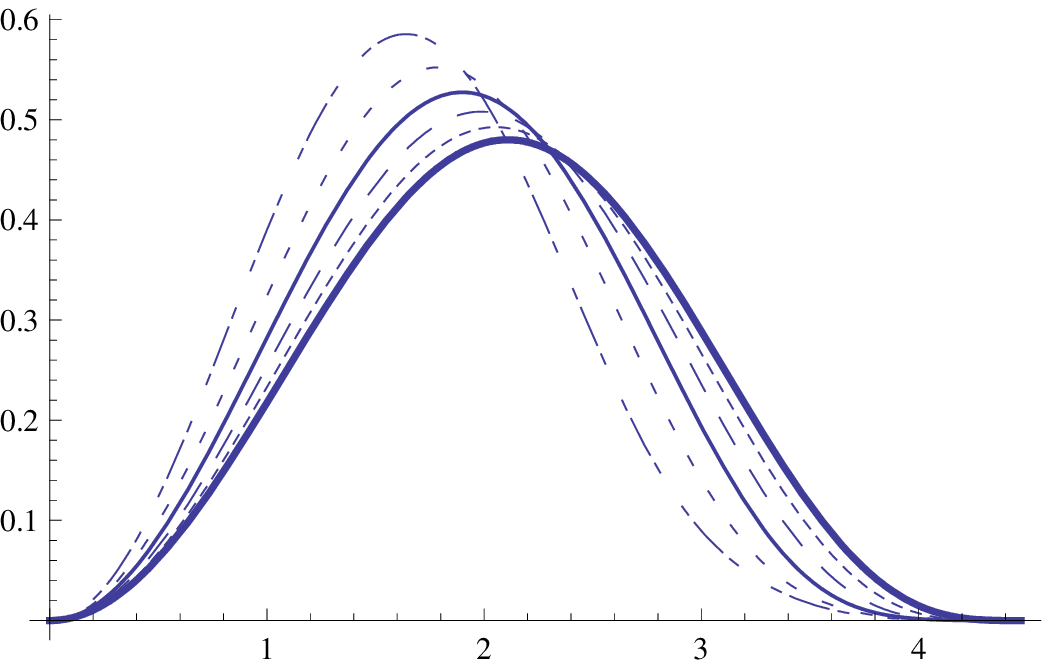}
\\
{\hspace{-0.0cm} {$y^{\prime}\longrightarrow $}}\\
 \caption{We show the angular average of the properly normalized velocity distribution  for values of the asymmetry parameter $\beta=(0.0.1,0.2,0.3, 0.4,0.5)$. In the plots $\beta$  is increasing from right to left). The results depend on the value of the potential in our vicinity. Here the value of $\xi(xs)=10$ was adopted. 
 %The similarity with Fig. \ref{Fig:toyfv1}(b) should be noted.
 }
 \label{Fig:toyfv2}
  \end{center}
  \end{figure}

   The asymmetric velocity  distribution in the galactic frame can be  written as:
   \beq
g(\beta,y^{\prime})=\frac{1}{1-\beta}f_{0,0,\infty}\left (\Phi(r_s)+\frac{1}{2} \left(\frac{1}{1-\beta}(y^{\prime 2}-\beta y_r ^{\prime 2}) \right )\right )
   \eeq
   This function depends on two variables. In order to compare with the previous results we exhibit in Fig.  \ref{Fig:toyfv2}(b) the dependence  on the asymmetry of its angular average. The values of $\beta$ employed were related to $a$ as above. We intend, however, to treat $\beta$ as a free parameter. The  results shown here exhibit the same trends as those obtained by using, e.g,  Tsallis functions (see \cite{VerHanH}). 
  
   Our next task is to transform the velocity distribution from the galactic to the local frame. The needed equation, see e.g. \cite{Vergados12}, is:
   \beq
{\bf y} \rightarrow {\bf y}+{\hat\upsilon}_s+\delta \left (\sin{\alpha}{\hat x}-\cos{\alpha}\cos{\gamma}{\hat y}+\cos{\alpha}\sin{\gamma} {\hat \upsilon}_s\right ) ,\quad y=\frac{\upsilon}{\upsilon_0}
\label{Eq:vlocal}
\eeq
with $\gamma\approx \pi/6$, $ {\hat \upsilon}_s$ a unit vector in the Sun's direction of motion, $\hat{x}$  a unit vector radially out of the galaxy in our position and  $\hat{y}={\hat \upsilon}_s\times \hat{x}$. The last term in the first expression of Eq. (\ref{Eq:vlocal}) corresponds to the motion of the Earth around the Sun with $\delta$ being the ratio of the modulus of the Earth's velocity around the Sun divided by the Sun's velocity around the center of the Galaxy, i.e.  $\upsilon_0\approx 220$km/s and $\delta\approx0.135$. The above formula assumes that the motion  of both the Sun around the Galaxy and of the Earth around the Sun are uniformly circular. The exact orbits are, of course, more complicated \cite{GREEN04}, but such deviations are not expected to significantly modify our results. In Eq. (\ref{Eq:vlocal}) $\alpha$ is the phase of the Earth ($\alpha=0$ around June 3nd)\footnote{One could, of course, make the time dependence of the rates due to the motion of the Earth more explicit by writing $\alpha \approx(6/5)\pi\left (2 (t/T)-1 \right )$, where $t/T$ is the fraction of the year.}.
\subsection{Standard non directional experiments}
We have seen that in the galactic frame in the presence of asymmetry $\beta$ the relevant quantity is:
$$ y^{\prime 2}_x+\frac{1}{1-\beta}\left (y^{\prime 2}_y+y^{\prime 2}_z\right )=\frac{1}{1-\beta}\left(y^{\prime 2}-\beta y^{\prime 2}_x  \right )$$
In the local frame the components $y_x,y_y,y_z$ of the velocity vector ${\bf y}$ are thus given by:
\barr
y_r&=&y_x=\frac{1}{sc}(y \cos \phi  \sin \theta +\delta  \sin \alpha  ),\quad
y_t=\sqrt{y^2_y+y_z^2}\nonumber\\
y_y&=&\frac{1}{sc}( y \sin \theta  \sin \phi -\delta  \cos \alpha  \cos \gamma ),\quad
y_z=\frac{1}{sc}(y \cos \theta +\delta  \cos \alpha  \sin \gamma +1),\nonumber\\
y&=&\frac{\upsilon}{\upsilon_0}
%sc=\frac{\sqrt{2 |\Phi(x_s)|}}{\upsilon_{\mbox{rot}}(x_s)}
\earr
where where $s_c$ is a suitable scale factor to bring the WIMP velocity  into units of the sun's velocity, $y=\upsilon/\upsilon_0$, i.e. $sc=\sqrt{|\Phi_0|}/\upsilon_0$.
One finds
\barr
&&\frac{1}{1-\beta}\left(y^{\prime 2}-\beta y^{\prime 2}_x  \right )\rightarrow Y^2=\frac{1}{sc^2}\frac{1}{1-\beta }\left(-\beta(\delta  \sin (\alpha )+y \cos (\phi ) \sin (\theta ))^2+\right . \nonumber\\
&&\left . (y \cos (\theta
   )+\delta  \cos (\alpha ) \sin (\gamma )+1)^2+(\delta  \cos (\alpha ) \cos (\gamma
   )-y \sin (\theta ) \sin (\phi ))^2\right )\nonumber\\
\earr
Thus the velocity distribution for the standard (non directional) case becomes:
\beq
g_{\mbox{{\tiny nodir}}}(Y)=\frac{1}{1-\beta}f_{0,0,\infty}\left (\Phi(r_s)+\frac{1}{2} Y^2\right )
\eeq
   \subsection{directional experiments}
 
In the Eddington theory the asymmetric velocity distribution is given by:
\beq
g_{\mbox{{\tiny dir}}}(X)=\frac{1}{1-\beta}f_{0,0,\infty}\left (\Phi(r_s)+\frac{1}{2}X^2\right )
\eeq
where $f$ is the symmetric normalized velocity distribution with respect to the center of the galaxy, $\beta$ is the asymmetry parameter and 
%$X=\frac{1}{1-\beta}\frac{1}{\upsilon_m}\left (\sqrt{\upsilon^2-\beta \upsilon_x^2} \right )$, $\upsilon_m$ the maximum allowed velocity. In the local frame
 X is given, \cite{VerMou11}, by:
\barr
&&X^2=\frac{1}{(1-\beta)s_c^2}\nonumber\\
&&
\left(\sqrt{3} \delta  \cos {\alpha } \cos {\Phi }-2 \sqrt{1-\xi^2}
   \sin {\phi }+2 \delta  \sin {\alpha } \sin {\Phi }\right)^2-\nonumber\\
   &&\beta 
   \left(2 \sqrt{1-\xi^2} \cos {\phi }-(\delta  \cos {\alpha }+2) \sin
   {\Theta }+\right .\nonumber\\ 
   && \left . \delta  \cos {\Theta } \left(2 \cos {\Phi } \sin {\alpha
   }-\sqrt{3} \cos {\alpha } \sin {\Phi }\right)\right)^2+\nonumber\\
   &&\left(2
   \xi y+(\delta  \cos {\alpha }+2) \cos {\Theta }+\delta  \sin {\Theta }
   \left(2 \cos {\Phi } \sin {\alpha }-\sqrt{3} \cos {\alpha } \sin
   {\Phi }\right)\right)^2+\nonumber\\
   &&\left(-2 \sqrt{1-\xi^2} \cos {\phi
   }+(\delta  \cos {\alpha }+2) \sin {\Theta }+ \right .\nonumber\\
   && \left . \delta  \cos {\Theta }
   \left(\sqrt{3} \cos {\alpha } \sin {\Phi }-2 \cos {\Phi } \sin
   {\alpha }\right)\right)^2.
\earr
%where $s_c$ is a suitable scale factor to bring the WIMP velocity  into units of the sun's velocity, $y=\upsilon/\upsilon_0$. $x=a\sqrt{u}$, with $u$ the energy transfer in dimensionless units, $u=(1/2)q^2 b^2$.
 The direction of the WIMP velocity is specified by $\xi=\cos{\theta}$ and $\phi$. 
The direction of observation is specified by the angles $\Theta$ and $\Phi$.
\section{Discussion}
In the present work we studied how one can construct the velocity distribution in the Eddington approach starting from dark matter density profiles. By modifying the distribution function by suitable angular momentum functions one can obtain asymmetric velocity distributions as well. We clarified some of the issues involved in this approach by considering a simple model which can yield analytic solutions. 
Results of realistic calculations for dark matter searches, employing the present technique and using realistic density profiles \cite{NFW}, \cite{ULLIO}, \cite{VEROW06}, will appear elsewhere \cite{MouOwVer14}.
%\\We find $s_c$=1.6 and 1.2 for the NFW and VO densities respectively. This leads to maximum velocities $2.10\upsilon_0$ and $4.0 \upsilon_0$ respectively.
\\
 {\bf Acknowledgments}: This research  has been partially supported  by the European  Union Social Fund (ESF) and the Greek national funds through the
 %the Operational Program "Education and Lifelong Learning" of the National Strategic Reference Framework (NSRF) - Research Funding 
 program THALIS  of the Hrllenic Open University: Development and Applications of Novel Instrumentation and Experimental Methods in Astroparticle  Physics.
% \bibliographystyle{apj}
% \bibliography{Tex}

\begin{thebibliography}{}
\expandafter\ifx\csname natexlab\endcsname\relax\def\natexlab#1{#1}\fi

\bibitem[{Ade {et~al.}(2013)}]{PlanckCP13}
Ade, A. P.~R., {et~al.} 2013, arXiv:1303.5076 [astro-ph.CO];The Planck
  Collaboration

\bibitem[{Agsn {et~al.}(2011)Agsn, Pedraza, \& Ramos-Caro}]{AgPedRM11}
Agsn, C.~A., Pedraza, J.~F., \& Ramos-Caro, J. 2011, Phys. Rev D, 83, 123007,
  arXiv:1104.5262 [gr-qc]

\bibitem[{Arnowitt \& Nath(1995)}]{ref2b}
Arnowitt, R., \& Nath, P. 1995, Phys. Rev. Lett., 74, 4592

\bibitem[{Arnowitt \& Nath(1996)}]{ref2c}
---. 1996, Phys. Rev. D, 54, 2374, hep-ph/9902237

\bibitem[{Bennett \& {\it et al}(1995)}]{Benne}
Bennett, D.~P., \& {\it et al}. 1995, Phys. Rev. Lett., 74, 2867

\bibitem[{Binney \& Tremain(2008)}]{bookBT87}
Binney, J., \& Tremain, S. 2008, Galactic Dynamics (Princeton University Press,
  Princeton, NJ, USA)

\bibitem[{Bottino {et~al.}(1997)}]{ref2a}
Bottino, A., {et~al.} 1997, Phys. Lett B, 402, 113

\bibitem[{Cochanek(1996)}]{Kochanek95}
Cochanek, C.~S. 1996, Astrophys. J., 457, 228, arXiv:astro-ph/9505068

\bibitem[{Collar \& {\it et al}(1992)}]{Drukb}
Collar, J., \& {\it et al}. 1992, Phys. Lett B, 275, 181

\bibitem[{Copi {et~al.}(1999)Copi, Heo, \& Krauss}]{Krauss}
Copi, C., Heo, J., \& Krauss, L. 1999, Phys. Lett. B, 461, 43

\bibitem[{Drukier {et~al.}(1986)Drukier, Freeze, \& Spergel}]{Druka}
Drukier, A.~K., Freeze, K., \& Spergel, D.~N. 1986, Phys. Rev. D, 33, 3495

\bibitem[{Eddington(1916)}]{EDDIN}
Eddington, A.~S. 1916, NRAS, 76, 572

\bibitem[{Ellis \& Flores(1991)}]{ELLFLOR}
Ellis, J., \& Flores, R.~A. 1991, Phys. Lett. B, 263, 259, {\it Phys. Lett. B}
  {\bf 300}, 175 (1993); {\it Nucl. Phys. B} {\bf 400}, 25 (1993)

\bibitem[{Ellis \& Roszkowski(1992)}]{ELLROSZ}
Ellis, J., \& Roszkowski, L. 1992, Phys. Lett. B, 283, 252

\bibitem[{Evans {et~al.}(2000)Evans, Carollo, \& Zeeuw}]{evans00}
Evans, N., Carollo, M., \& Zeeuw, P. 2000, Mon.~Not.~R.~Astron.~Soc, 318, 1131

\bibitem[{Foot(2011)}]{Foot11}
Foot, R. 2011, Phys. Lett. B, 703, 7, [arXiv:1106.2688]

\bibitem[{Foot {et~al.}(1991)Foot, Lew, \& Volkas}]{FLV72}
Foot, R., Lew, H., \& Volkas, R.~R. 1991, Phys. Lett. B, 272, 676

\bibitem[{Fornasa \& Green(2013)}]{FornGreen13}
Fornasa, M., \& Green, A. 2013, arXiv:1311.5477 (astro-ph.CO]

\bibitem[{Gelmini \& Gondolo(2001)}]{Gelmini}
Gelmini, G., \& Gondolo, P. 2001, Phys. Rev. D, 64, 123504

\bibitem[{G\'{o}mez {et~al.}(2000)G\'{o}mez, Lazarides, \& Pallis}]{Gomez2}
G\'{o}mez, M.~E., Lazarides, G., \& Pallis, C. 2000, Phys. Rev. D, 61, 123512

\bibitem[{G\'{o}mez \& Vergados(2001)}]{Gomez1}
G\'{o}mez, M.~E., \& Vergados, J.~D. 2001, Phys. Lett. B, 512, 252, ;
  hep-ph/0012020

\bibitem[{Green(2003)}]{GREEN04}
Green, A. 2003, Phys. Rev. D, 68, 023004, ibid: D ${\bf 69}$ (2004) 109902;
  arXiv:astro-ph/0304446

\bibitem[{Green(2001)}]{Green}
Green, A.~M. 2001, Phys. Rev. D, 63, 103003

\bibitem[{Green(2002)}]{GREEN02}
---. 2002, Phys. Rev. D, 66, 083003

\bibitem[{Gudnason {et~al.}(2006)Gudnason, Kouvaris, \& Sannino}]{GKS06}
Gudnason, S.~B., Kouvaris, C., \& Sannino, F. 2006, Phys. Rev. D, 74, 095008,
  arXiv:hep-ph/0608055

\bibitem[{Halverson {et~al.}(2002)}]{DASI02}
Halverson, N.~W., {et~al.} 2002, Astrophys. J., 568, 38

\bibitem[{Hanary \& {\it et al}(2000)}]{MAXIMA1}
Hanary, S., \& {\it et al}. 2000, Astrophys. J., 545, L5

\bibitem[{Hansen {et~al.}(2006)Hansen, Moore, Zemp, \& Stadel}]{HANSEN06a}
Hansen, S.~H., Moore, B., Zemp, M., \& Stadel, J. 2006, JCAP, 0601, 014

\bibitem[{Jaffe \& {\it et al}(2001)}]{flat01}
Jaffe, A.~H., \& {\it et al}. 2001, Phys. Rev. Lett., 86, 3475

\bibitem[{Mauskopf \& {\it et al}(2002)}]{BOOMERANG1}
Mauskopf, P.~D., \& {\it et al}. 2002, Astrophys. J., 536, L59

\bibitem[{Merritt(1985{\natexlab{a}})}]{MERRITT}
Merritt, D. 1985{\natexlab{a}}, A J, 90, 1027

\bibitem[{Merritt(1985{\natexlab{b}})}]{MERRITT85b}
---. 1985{\natexlab{b}}, MNRAS, 214, 25

\bibitem[{Mosi \& {\it et al}(2002)}]{BOOMERANG2}
Mosi, S., \& {\it et al}. 2002, Prog. Nuc.Part. Phys., 48, 243

\bibitem[{Moustakidis {et~al.}(2014)Moustakidis, Owen, \&
  Vergados}]{MouOwVer14}
Moustakidis, C.~C., Owen, D., \& Vergados, J. 2014, (to be published)

\bibitem[{Navarro {et~al.}(1996)Navarro, Frenk, \& White}]{NFW}
Navarro, J., Frenk, C., \& White, S. 1996, ApJ, 462, 563

\bibitem[{Nguyen \& Pedraza(2013)}]{NguPed13}
Nguyen, P.~H., \& Pedraza, J.~F. 2013, Phys. Rev D, 88, 064020, arXiv:1305.7220
  [gr-qc]

\bibitem[{Nussinov(1992)}]{Nussinov92}
Nussinov, S. 1992, Phys. Lett. B, 279, 111

\bibitem[{Oikonomou {et~al.}(2007)Oikonomou, Vergados, \&
  Moustakidis}]{OikVerMou}
Oikonomou, V., Vergados, J., \& Moustakidis, C.~C. 2007, Nuc. Phys., B 773, 19

\bibitem[{Osipkov(1979)}]{OSIPKOV79}
Osipkov, L. 1979, Soviet Astronomy Letters, 5, 42

\bibitem[{Owen \& Vergados(2003)}]{OWVER}
Owen, D., \& Vergados, J.~D. 2003, Astrophys. J., 589, 17, astro-ph/0203923

\bibitem[{Plummer(1911)}]{PLUMMER}
Plummer, H.~C. 1911, MNRAS, 71, 460

\bibitem[{Santos \& {\it et al}(2002)}]{MAXIMA3}
Santos, M., \& {\it et al}. 2002, Phys. Rev. Lett., 88, 241302

\bibitem[{Servant \& Tait(2003)}]{ST02a}
Servant, G., \& Tait, T. M.~P. 2003, Nuc. Phys. B, 650, 391

\bibitem[{Sikivie(1998)}]{SIKIVI2}
Sikivie, P. 1998, Phys. Lett. B, 432, 139

\bibitem[{Sikivie(1999)}]{SIKIVI1}
---. 1999, Phys. Rev. D, 60, 063501

\bibitem[{Smoot \& {\it et al}~(COBE~Collaboration)(1992)}]{COBE}
Smoot, G.~F., \& {\it et al}~(COBE~Collaboration). 1992, Astrophys. J., 396, L1

\bibitem[{Spergel {et~al.}(2007)}]{WMAP06}
Spergel, D., {et~al.} 2007, Astrophys. J. Suppl., 170, 377,
  [arXiv:astro-ph/0603449v2]

\bibitem[{Spergel \& {\it et al}(2003)}]{SPERGEL}
Spergel, D.~N., \& {\it et al}. 2003, Astrophys. J. Suppl., 148, 175

\bibitem[{Ullio \& Kamioknowski(2001)}]{UK01}
Ullio, P., \& Kamioknowski, M. 2001, JHEP, 0103, 049

\bibitem[{Ullio \& Kamionkowski(2001)}]{ULLIO}
Ullio, P., \& Kamionkowski, M. 2001, JHEP, 0103, 049

\bibitem[{Vergados {et~al.}(2008)Vergados, Hansen, \& Host}]{VerHanH}
Vergados, J., Hansen, S.~N., \& Host, O. 2008, Phys. Rev. D, 77, 023509

\bibitem[{Vergados \& Owen(2007)}]{VEROW06}
Vergados, J., \& Owen, D. 2007, Phys. Rev., D 75, 043503

\bibitem[{Vergados(2000)}]{Verg00}
Vergados, J.~D. 2000, Phys. Rev. D, 62, 023519

\bibitem[{Vergados(2001)}]{Verg01}
---. 2001, Phys. Rev. D, 63, 06351

\bibitem[{Vergados(2007)}]{JDV06a}
---. 2007, Lect. Notes Phys., 720, 69, hep-ph/0601064

\bibitem[{Vergados(2012)}]{Vergados12}
---. 2012, Phys. Rev D, 85, 123502, arXiv:1202.3105 (hep-ph)

\bibitem[{Vergados \& Moustakidis(2011)}]{VerMou11}
Vergados, J.~D., \& Moustakidis, C.~C. 2011, Eur. J. Phys., 9(3), 628,
  arXiv:0912.3121 [astro-ph.CO]

\bibitem[{Wojtak {et~al.}(2008)Wojtak, Lokas, Mamon, Gottloeber, Klypin, \&
  Hoffman}]{Asym08}
Wojtak, R., Lokas, E.~L., Mamon, G.~A., {et~al.} 2008, Mon. Not. Roy. Astron.
  Soc., 388, 815, arXiv:0802.0429 (astro-ph)

\bibitem[{Wu \& {\it et al}(2001)}]{MAXIMA2}
Wu, J., \& {\it et al}. 2001, Phys. Rev. Lett., 87, 251303

\end{thebibliography}

 \end{document}